\documentclass[]{spie}  
\usepackage{multicol}
\usepackage{gensymb}
\usepackage{float}
\usepackage{array}

\usepackage[encapsulated]{CJK}
\usepackage{ucs}
\newcommand{\cntext}[1]{\begin{CJK}{UTF8}{gbsn}#1\end{CJK}}

\usepackage{amsmath,amsfonts,amssymb}
\usepackage{graphicx,subcaption}
\usepackage[colorlinks=true, allcolors=blue]{hyperref}
\usepackage{orcidlink}

\title{Design and characterization of new 90~GHz detectors for the Cosmology Large Angular Scale Surveyor (CLASS)}

\author[a]{Carolina N\'u\~nez\orcidlink{0000-0002-5247-2523}}
\author[a]{John~W. Appel\orcidlink{0000-0002-8412-630X}}
\author[a]{Sarah Marie Bruno\orcidlink{0000-0003-2682-7498}}
\author[a]{Rahul Datta\orcidlink{0000-0003-3853-8757}}
\author[b,a]{Aamir Ali\orcidlink{0000-000
1-7941-9602}}
\author[a]{Charles~L. Bennett\orcidlink{0000-0001-8839-7206}}
\author[c,a]{Sumit Dahal\orcidlink{0000-0002-1708-5464}}
\author[a]{Jullianna Denes~Couto\orcidlink{0000-0002-0552-3754}}
\author[c]{Kevin~L. Denis\orcidlink{0000-0002-3592-5703}}
\author[a]{Joseph Eimer\orcidlink{0000-0001-6976-180X}}
\author[d]{Francisco Espinoza\orcidlink{0000-0002-1052-0339}}
\author[c,a]{Tom Essinger-Hileman\orcidlink{0000-0002-4782-3851}}
\author[e,c]{Kyle Helson\orcidlink{0000-0001-9238-4918}}
\author[f,a]{Jeffrey Iuliano\orcidlink{0000-0001-7466-0317}}
\author[a]{Tobias~A. Marriage\orcidlink{0000-0003-4496-6520}}
\author[a]{Carolina Morales P\'erez\orcidlink{0000-0002-1371-5334}}
\author[a]{Deniz Augusto~Nunes~Valle\orcidlink{0000-0003-3487-2811}}
\author[g,a]{Matthew A. Petroff\orcidlink{0000-0002-4436-4215}}
\author[c]{Karwan Rostem\orcidlink{0000-0003-4189-0700}}
\author[a]{Rui Shi~(\cntext{时瑞})\orcidlink{0000-0001-7458-6946}}
\author[h,a]{Duncan~J. Watts\orcidlink{0000-0002-5437-6121}}
\author[c]{Edward~J. Wollack\orcidlink{0000-0002-7567-4451}}
\author[i,f,a]{Zhilei Xu~(\cntext{徐智磊})\orcidlink{0000-0001-5112-2567}}

\affil[a]{The William H. Miller III Department of Physics and Astronomy, Johns Hopkins University, 3701 San Martin Drive, Baltimore, MD 21218, USA}

\affil[b]{Department of Physics, University Of California, Berkeley, CA 94720, USA}
\affil[c]{Goddard Space Flight Center, 8800 Greenbelt Road, Greenbelt, MD 20771, USA}
\affil[d]{Facultad de Ingenier\'ia, Universidad Cat\'olica de la Sant\'isima Concepci\'on, Alonso de Ribera 2850, Concepci\'on, Chile}
\affil[e]{Center for Space Sciences and Technology, University of Maryland, Baltimore County, 1000 Hilltop Circle, Baltimore, MD 21250}
\affil[f]{Department of Physics and Astronomy, University of Pennsylvania, 209 South 33rd Street, Philadelphia, PA 19104, USA}
\affil[g]{Center for Astrophysics, Harvard \& Smithsonian, 60 Garden Street, Cambridge, MA 02138, USA}
\affil[h]{Institute of Theoretical Astrophysics, University of Oslo, P.O.Box 1029 Blindern, N-0315 Oslo, Norway}
\affil[i]{MIT Kavli Institute, Massachusetts Institute of Technology, 77 Massachusetts Avenue, Cambridge, MA 02139, USA}

\authorinfo{Further author information: (Send correspondence to C. N.)\\C. N.: E-mail:cnunez4@jhu.edu}

\begin{document} 
\maketitle

\begin{abstract}
The Cosmology Large Angular Scale Surveyor (CLASS) is a polarization-sensitive telescope array located at an altitude of 5,200 m in the Chilean Atacama Desert.
CLASS is designed to measure ``E-mode'' (even parity) and ``B-mode'' (odd parity) polarization patterns in the Cosmic Microwave Background (CMB) over large angular scales with the aim of improving our understanding of inflation, reionization, and dark matter.
CLASS is currently observing with three telescopes covering four frequency bands: one at 40~GHz (Q); one at 90~GHz (W1); and one dichroic system at 150/220~GHz (G).
In these proceedings, we discuss the updated design and in-lab characterization of new 90~GHz detectors.
The new detectors include design changes to the transition-edge sensor (TES) bolometer architecture, which aim to improve stability and optical efficiency.
We assembled and tested four new detector wafers, to replace four modules of the W1 focal plane.
These detectors were installed into the W1 telescope, and will achieve first light in the austral winter of 2022.
We present electrothermal parameters and bandpass measurements from in-lab dark and optical testing.
From in-lab dark tests, we also measure a median NEP of 12.3 $\mathrm{aW\sqrt{s}}$ across all four wafers about the CLASS signal band, which is below the expected photon NEP of 32 $\mathrm{aW\sqrt{s}}$ from the field.  We therefore expect the new detectors to be photon noise limited.

\end{abstract}


\section{INTRODUCTION}

\label{sec:introduction}  

The Cosmology Large Angular Scale Surveyor (CLASS) is a polarization-sensitive telescope array located at an altitude of 5,200 m in the Chilean Atacama Desert. 
CLASS is designed to measure ``E-mode'' (even parity) and ``B-mode'' (odd parity) polarization patterns in the Cosmic Microwave Background (CMB) over large angular scales ($>1\degree$). 
CLASS seeks to improve our understanding of inflation, reionization, and dark matter \cite{Tom-overview, Harrington-overview}. 

CLASS is currently observing with three telescopes covering four frequency bands: one at 40~GHz (Q); one at 90~GHz (W1); and one dichroic system at 150/220~GHz (G).
The Q, W1, and G telescopes have been observing since June 2016, May 2018, and September 2019, respectively.
In-lab characterization and on-sky performance of these detector arrays has been discussed in Refs.~\citenum{Appel-Q, Dahal-W1, Dahal-HF, Dahal-multifrequency}.
In these proceedings, we discuss the redesign and preliminary in-lab characterization of new 90~GHz detectors.
Design changes were made with the goal of improving detector stability and optical efficiency.
These detectors are currently being fielded to upgrade the W1 focal plane.
We anticipate first light for the new W1 focal plane array in the austral winter of 2022.

\section{Focal Plane Design}
\label{sec:focal_plane_design}
   
\subsection{Detector Design}
\label{subsec:detector_design}

   \begin{figure} [ht]
   \begin{center}
   \includegraphics[width=\textwidth, trim={2cm 0 2cm 0}]{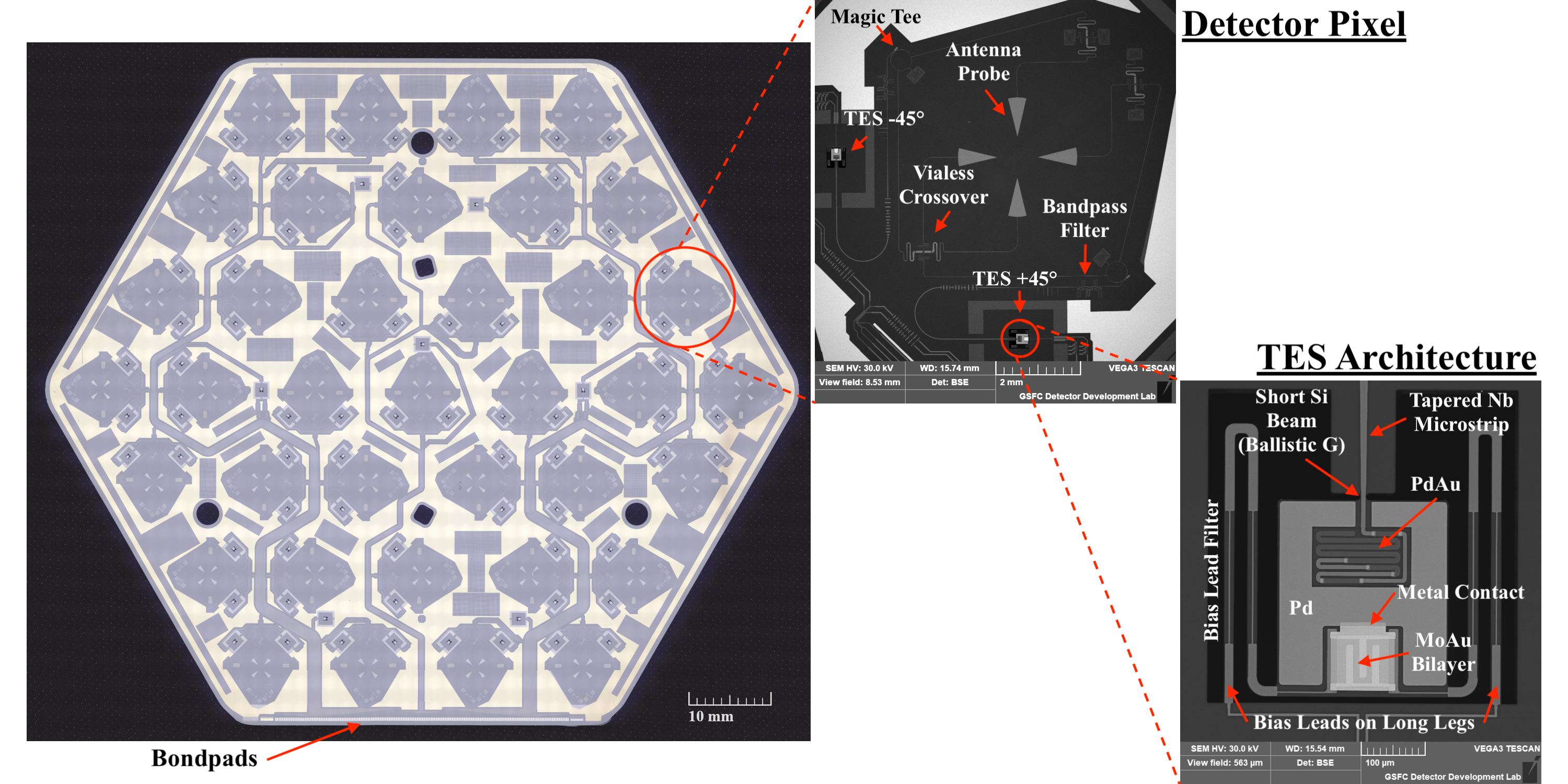}
   \end{center}
   \caption{ \label{fig:design} 
 CLASS 90~GHz wafer (left), with zoom in of a detector pixel and updated transition-edge sensor (TES) architecture. For a full description of the 90~GHz wafer design, see Refs.~\citenum{Chuss-development, Denis-fabrication, Rostem-design}. Each of the 37 detector pixels consists of a symmetric planar ortho-mode transducer (OMT), which reads out two orthogonal linear polarizations over Nb microstrip transmission lines to MoAu bilayer TES bolometers.
The updated TES architecture includes a simplified absorber with a resistive meander, a direct metal connection between the TES and the Pd, and a revised choke filter circuit implementation.
These changes were made to improve stability and optical efficiency performance of the TES bolometers.}
   \end{figure}

CLASS focal planes consist of arrays of highly sensitive feedhorn-coupled transition-edge sensor (TES) bolometers. 
TES bolometers provide background-limited sensitivity, required to achieve CLASS' science goals.
Their design is scalable to large focal plane arrays across multiple frequencies, required for high sensitivity measurements and separation of the CMB signal from polarized Galactic foregrounds.

In Figure~\ref{fig:design}, an example CLASS 90~GHz detector wafer is shown.
The detector wafers, which integrate 37 detector pixels, are fabricated at NASA Goddard Space Flight Center.
Following the blue leak mitigation strategy presented in Ref.~\citenum{Wollack-blueleak}, the detector wafer is indium-bump bonded to a backshort wafer for signal termination and a photonic choke~\cite{Crowe-choke} wafer, which serves as an interface to the sensor array’s feedhorns.
Each of the 37 detector pixels consists of a symmetric planar ortho-mode transducer (OMT), which reads out two orthogonal linear polarizations ($+45\degree$ and $-45\degree$ from the vertical) to two TES bolometers.
Signals from opposing antenna probes are coherently combined onto a single microstrip transmission line using the difference output of a Magic Tee, which transmits a single mode.~\cite{U-Yen-magicT}
On-chip filtering and micromachined silicon packaging define the signal bandpass.~\cite{Crowe-choke}
Finally, the signal from each of the two orthogonal polarizations is passed to a MoAu bilayer TES bolometer.
During operation, the TES bolometers are voltage-biased to their superconducting transition critical temperature ($T_\mathrm{c}$) of $\sim$150~mK.
For a full description of the original 90~GHz detector design, see Refs.~\citenum{Chuss-development, Denis-fabrication, Rostem-design}.

The updated TES architecture, shown in the rightmost panel of Figure~\ref{fig:design}, includes three primary design changes from the original CLASS 90~GHz detectors:
\begin{enumerate}
    \item a simplified absorber that terminates power from the sky (brought in via a Nb microstrip) onto the TES island, with a resistive PdAu meander that consists of a stepped impedance transition from Nb to PdAu; 
    \item the addition of a direct normal-metal connection between the TES and the heat capacity element formed by the Pd, to effectively lump the electronic heat capacity into a single element; 
    \item the revision of the choke filter circuit design to extend onto the membrane’s diffusive bolometer legs.
\end{enumerate}

These design changes were introduced in order to improve optical efficiency (1, 3), and improve stability of the TES transition (2), which are described in Ref.~\citenum{Dahal-multifrequency}.
In addition, the redesign introduced a more compact absorber at the Magic Tee and at the microstrip crossovers.

\subsection{Module Assembly Structure}
\label{subsec:module_assembly_structure}

\begin{figure} [ht!]
\begin{center}
   
\begin{subfigure}{0.45\textwidth}
   \includegraphics[width=\linewidth]{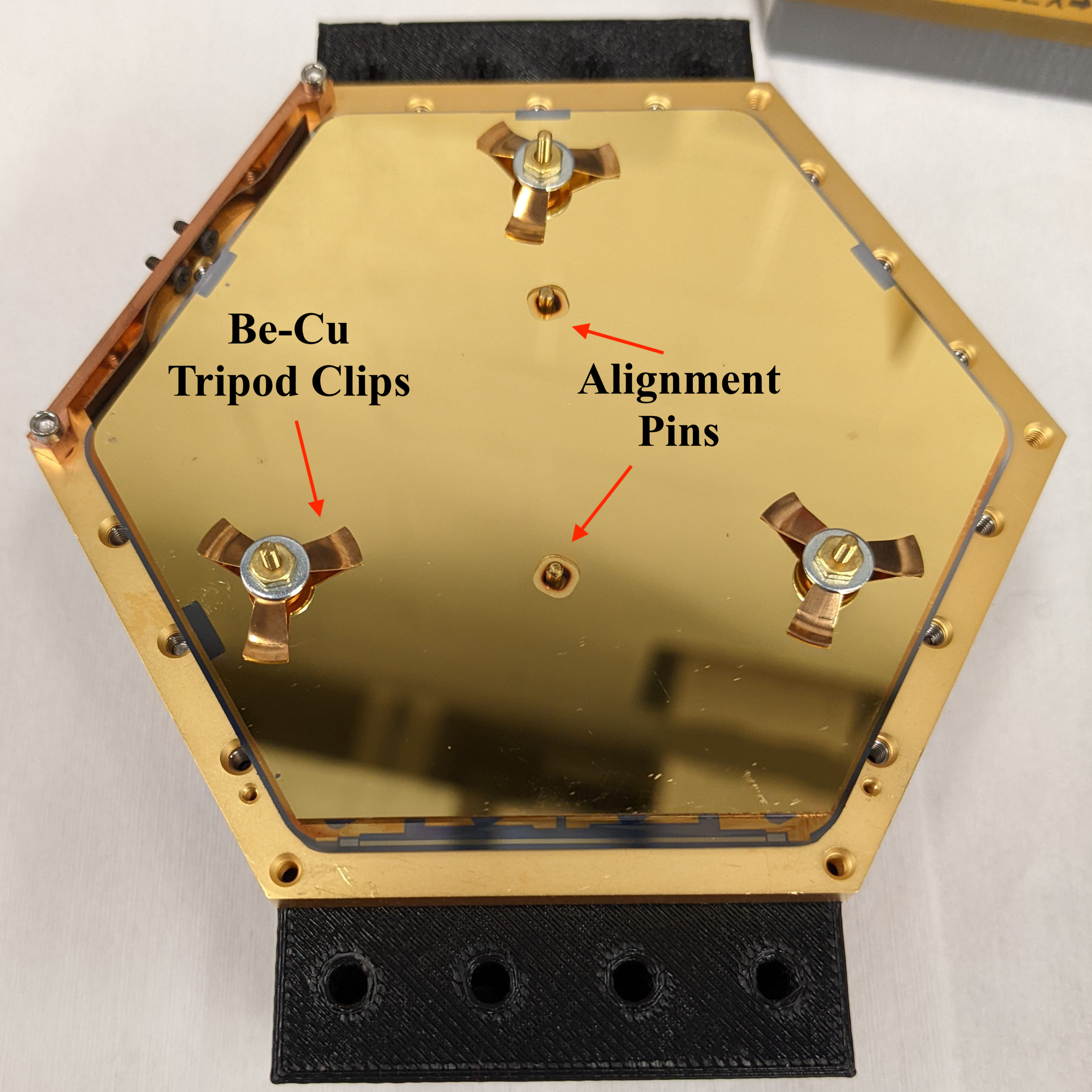}
   \caption{} \label{fig:mounted_wafer}
\end{subfigure} 
\hfill
\begin{subfigure}{0.45\textwidth}
   \includegraphics[width=\linewidth]{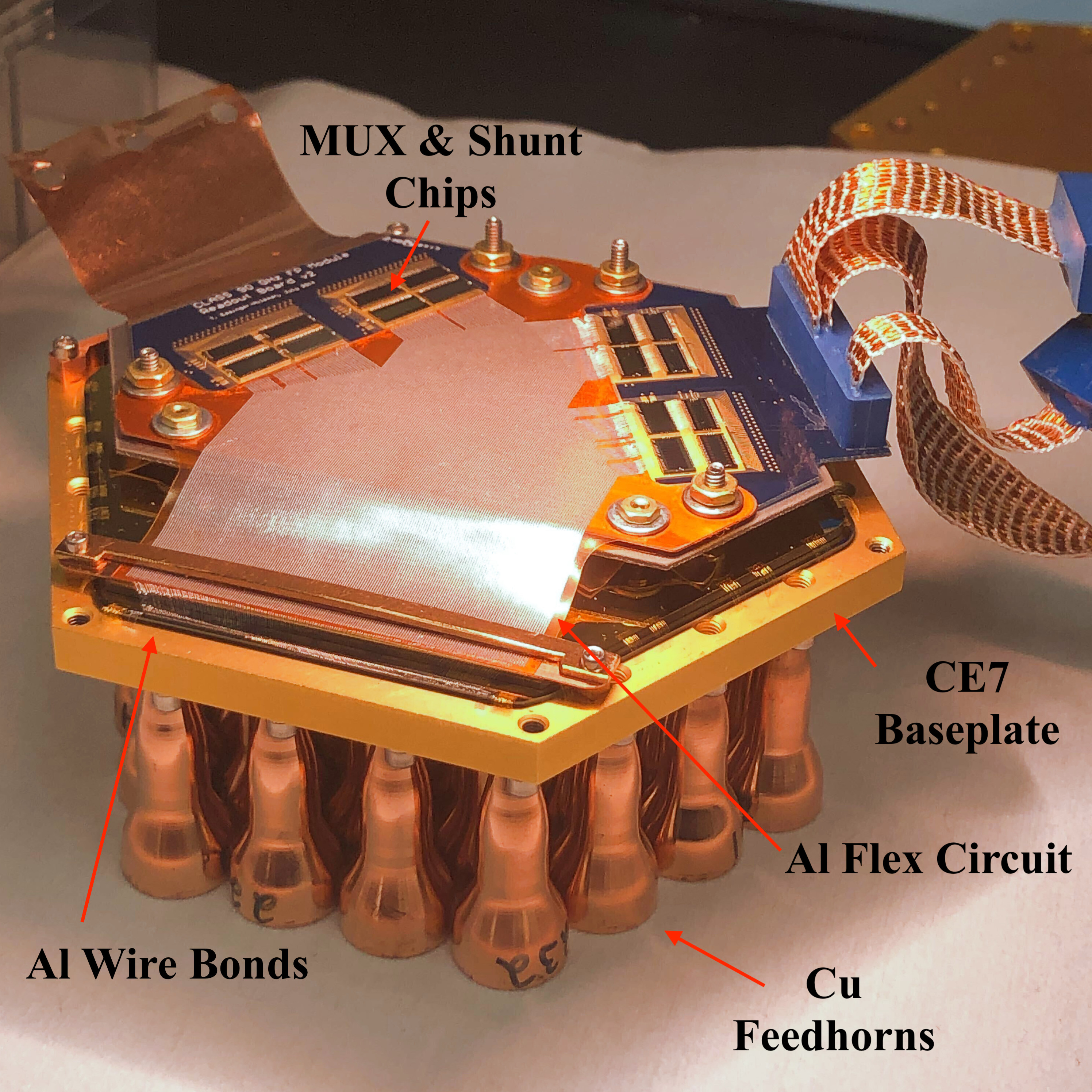}
   \caption{} \label{fig:bonded_module}
\end{subfigure}

\bigskip
\begin{subfigure}{0.45\textwidth}
   \includegraphics[width=\linewidth]{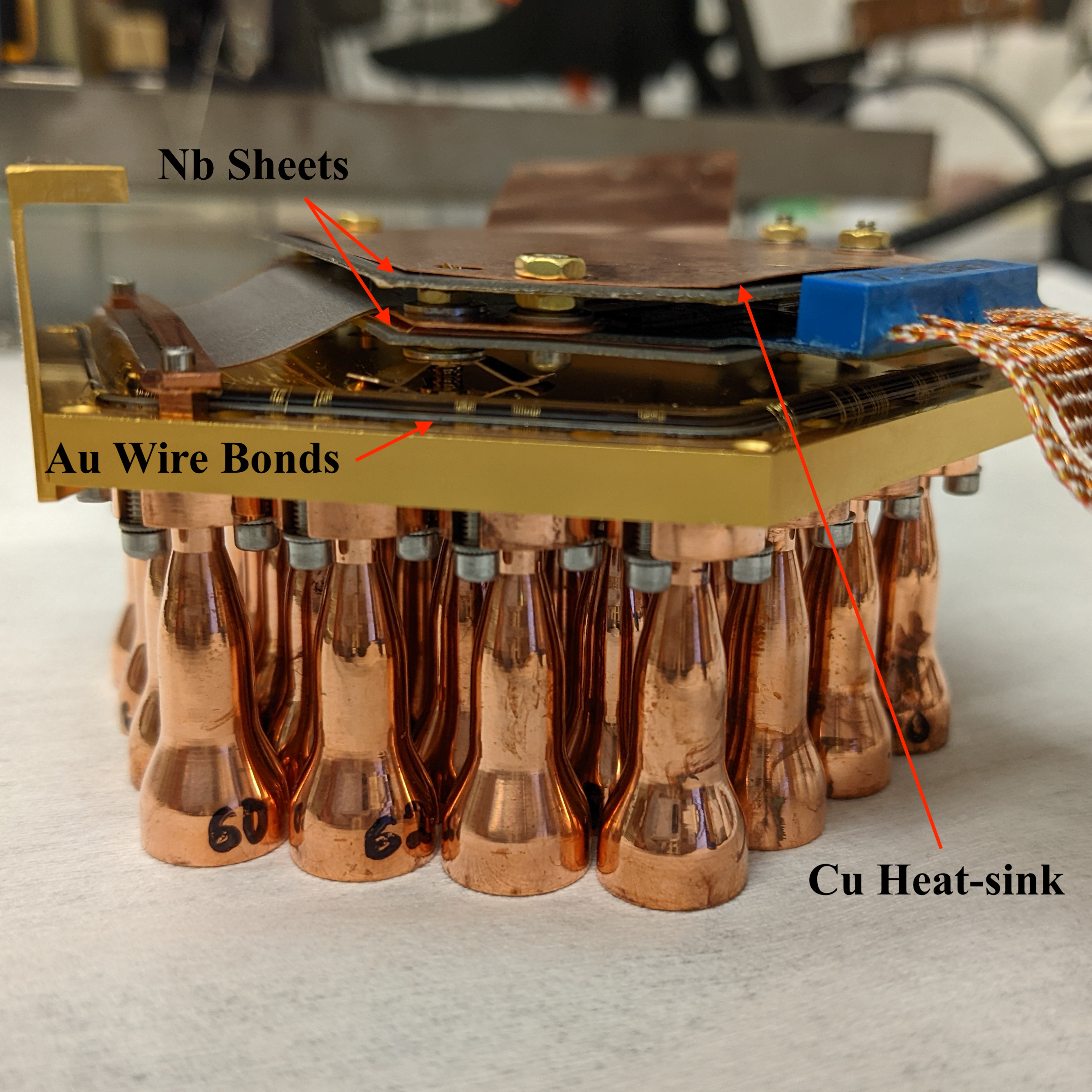}
   \caption{} \label{fig:side_view_assembly}
\end{subfigure}
\hfill
\begin{subfigure}{0.45\textwidth}
   \includegraphics[width=\linewidth]{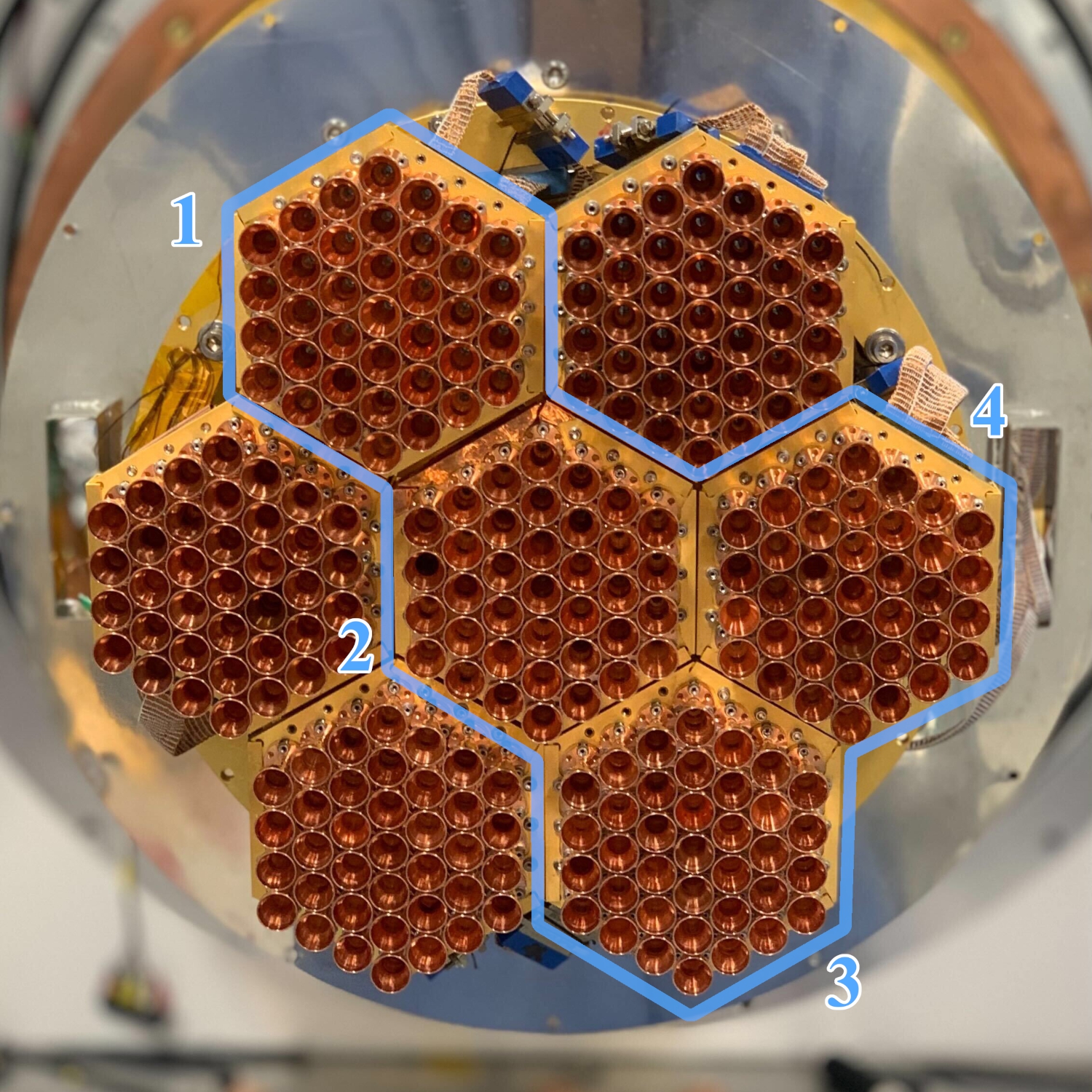}
   \caption{} \label{fig:W1_in_field}
\end{subfigure}
   
   \end{center}
   \caption[example] 
   { \label{fig:assembly} The CLASS detector assemblies are integrated into focal plane array modules for ease of testing and readout; a full focal plane consists of seven modules. Panel~\ref{fig:mounted_wafer} shows a CLASS wafer mounted and aligned onto a CE7 baseplate.  Panels~\ref{fig:bonded_module} and \ref{fig:side_view_assembly} show all interior layers of a fully integrated module, before it is packaged into its final configuration. For a full description of the assembly stack, see Ref.~\citenum{Dahal-W1}. Panel~\ref{fig:W1_in_field} shows the four new CLASS 90~GHz modules (outlined in blue) situated in the fielded W1 receiver. We anticipate first light for these new detectors in the austral winter of 2022.}
   \end{figure} 
  
The micromachined CLASS detector assemblies are subsequently integrated into focal plane array modules at Johns Hopkins University (JHU) to facilitate testing and readout.
A full 90~GHz focal plane consists of seven modules.
Figure~\ref{fig:assembly} highlights various stages of the assembly process, as well as the full upgraded W1 focal plane.
Smooth-walled feedhorns\cite{Feedhorns} couple light from the sky onto the TES bolometers.
The machined Cu feedhorns are individually installed onto the front of a Au-plated Si Alloy Controlled Expansion 7 (CE7) baseplate.\cite{CE7}
CE7, composed of  70\% Si and 30\% Al, is chosen due to its machinability and its low differential thermal contraction relative to Si.
The Si wafer is then mounted and aligned onto the baseplate using 1) two alignment pins and a Cu spring clip along one edge of the wafer to align the detector OMTs to the CE7 waveguide, and 2) three Be-Cu tripod spring clips to hold the Si wafer onto the baseplate.
Au wire bonds provide heat-sinking from the wafer to the baseplate, and Al bonding connects the detectors to the readout circuit (RC).
The RC consists of a printed circuit board (PCB), 8 shunt chips, 8 NIST-provided multiplexing (MUX) chips housing SQUIDs (Superconducting Quantum Interference Devices) for time-division multiplexed (TDM) readout, an Al flex circuit, Al wire bonds, and woven NbTi cable assemblies provided by Tekdata Interconnections, Ltd.\footnote{\url{https://www.tekdata-interconnect.com/}}
The RC is sandwiched between two Nb sheets, which provide magnetic shielding for the SQUIDs, and is heat-sunk with a Cu layer.
For a full description of the assembly stack, see Ref.~\citenum{Dahal-W1}.

\section{Detector Characterization}
\label{sec:detector_characterization}

In-lab characterization of the CLASS detectors is performed at JHU.
The CLASS detectors are mounted onto the 100~mK stage of a pulse tube pre-cooled dilution refrigerator \cite{Iuliano-receiver}.
The cryostat reaches an operational bath temperature ($T_{\mathrm{bath}}$) of $\sim$50~mK.
NbTi Tekdata wiring connects the feedback and bias lines from the focal plane to the SQUID Series Array (SSA) board at 4~K; the SSA board amplifies the signal to be read out at room temperature by the TDM Multi-Channel Electronics (MCE).\cite{TDM}

For estimation of electrothermal parameters and dark noise-equivalent power (NEP) was conducted in a dark configuration, with all stages of the cryostat closed with metal plates.
For bandpass measurements, the cryostat was placed in a necked-down optical configuration to prevent detector saturation, with a nylon filter on the 1~K stage, a small aperture plate with an anti-reflection coated nylon filter at the 4~K stage, and a small aperture plate with a polytetrafluoroethylene (PTFE) filter at the 60~K stage.
An ultra-high-molecular-weight polyethylene (UHMWPE) vacuum window lets in light at the front of the cryostat, while extruded polystyrene foam (XPS) filters at the 4~K stage rejects out-of-band infrared radiation.
Further in-lab characterization and on-sky performance of the new 90~GHz detectors will be discussed in a future publication.

\subsection{Electrothermal Parameters}
\label{subsec:electrothermal_parameters}

   \begin{figure} [ht]
   \begin{center}
   \includegraphics[width=\textwidth]{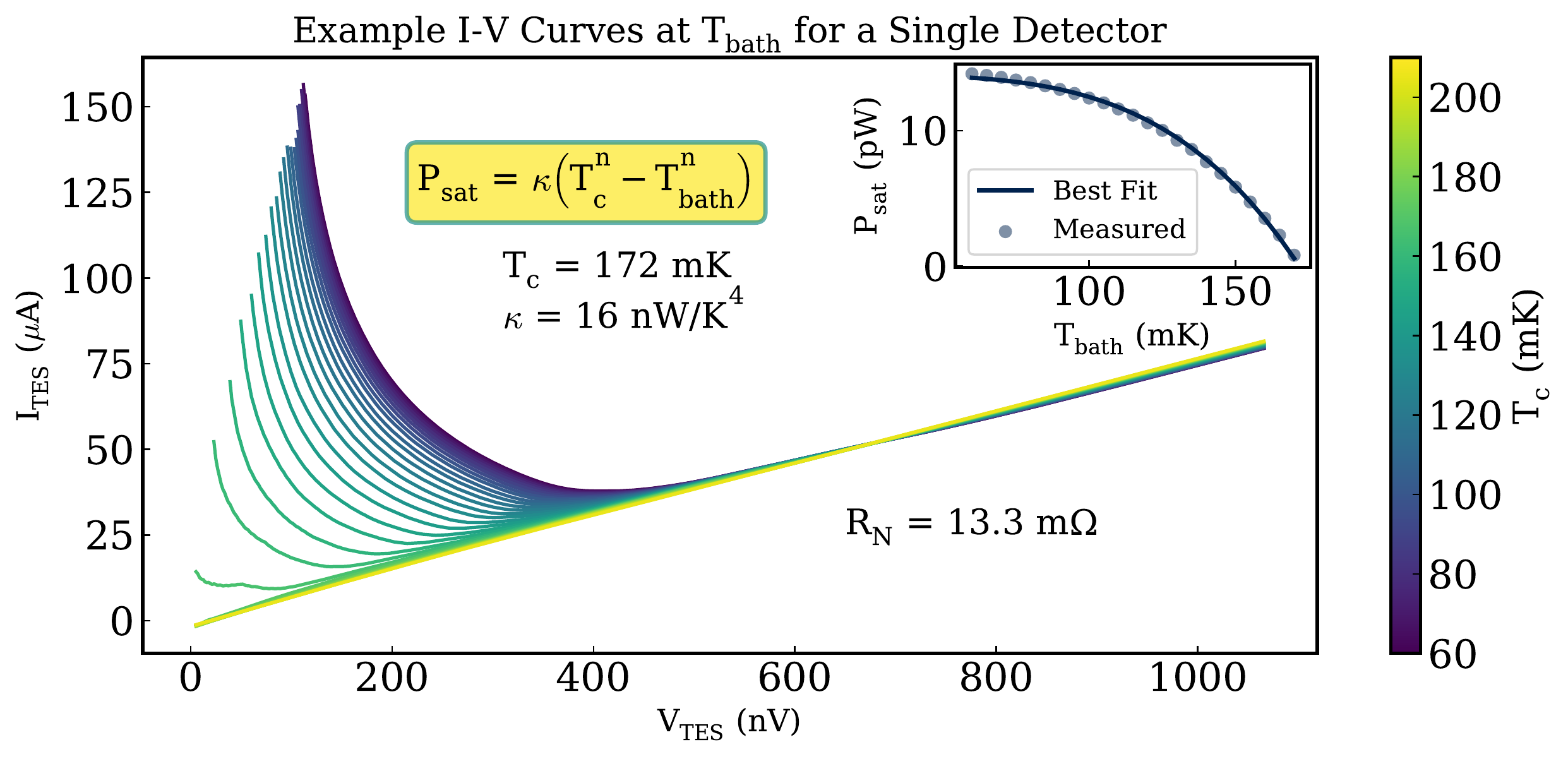}
   \end{center}
   \caption{ \label{fig:IVs} 
 Example I-V curves taken at multiple bath temperatures ($T_{\mathrm{bath}}$) for one detector. The inverse slope of the normal branch of the I-V yields the normal resistance ($R_\mathrm{N}$) of the TES. The saturation power ($P_{\mathrm{sat}}$) is defined to be $I_{\mathrm{TES}}\times V_{\mathrm{TES}}$ at 70\%~$R_\mathrm{N}$. The inset figure shows the measured $P_{\mathrm{sat}}$ at each $T_{\mathrm{bath}}$, from which the TES critical temperature, $T_\mathrm{c}$, and $\kappa$ can be determined.
 We assume $n=4$ (ballistic phonon limit).\cite{Appel-calibration, Dahal-W1, TES-chapter}}
   \end{figure}

I-V curves are used to determine the electrothermal characteristics of the TES bolometers. Over a wide range ($\sim$60-210~mK) of bath temperatures ($T_{\mathrm{bath}}$), we measure each I-V curve by ramping down the voltage bias in steps, driving the TES from normal to superconducting. We convert from the MCE  digital-to-analog (DAC) feedback units and voltage bias to TES current ($I_{\mathrm{TES}}$) and TES voltage ($V_{\mathrm{TES}}$) following \S~4.1 of Ref.~\citenum{Appel-calibration}.

Figure~\ref{fig:IVs} shows an example of this process for one detector.
The inverse slope of the normal branch of the I-V yields the normal resistance ($R_\mathrm{N}$) of the TES.
At each $T_{\mathrm{bath}}$, we measure the saturation power $P_{\mathrm{sat}}$, which is the amount of power, given by $I_{\mathrm{TES}}\times V_{\mathrm{TES}}$, required to maintain the TES at its superconducting critical temperature ($T_\mathrm{c}$) with a resistance equal to 70\% of the normal resistance $R_\mathrm{N}$\cite{Appel-calibration}.  We can then solve for $T_\mathrm{c}$ and $\kappa$ using the relation\cite{TES-chapter}:

\begin{equation}
    P_{\mathrm{sat}} = \kappa \left(T_\mathrm{c}^n - T_{\mathrm{bath}}^n\right) \mathrm{.}
\end{equation}

\noindent The normalization prefactor $\kappa$ is related to the geometry of the stubby ballistic beam\cite{Ballistic} shown in Figure~\ref{fig:design}. We assume $n=4$ (ballistic phonon limit). The thermal conductance ($G$) of the TES, also associated with the stubby ballistic beam, is given by\cite{TES-chapter}:

\begin{equation}
    G = \frac{dP}{dT}\Bigr|_{\substack{T_\mathrm{c}}} = n \kappa T_\mathrm{c}^{n-1} \mathrm{.}
\end{equation}

\begin{table}[ht]
\caption{Median (standard deviation) values for each wafer of key electrothermal parameters derived from I-V curves. We report values for approximately 45, 51, 47, and 32 bolometers for wafers 1--4, respectively.} 
\label{tab:parameters}
\begin{center}       
\begin{tabular}{|l|c|c|c|c|} 
\hline
\rule[-1ex]{0pt}{3.5ex}  & Wafer 1 & Wafer 2 & Wafer 3 & Wafer 4\\
\hline
\rule[-1ex]{0pt}{3.5ex}  $G$~(pW/$K$) @ $T_\mathrm{c}$ & 269 (63) & 229 (30) & 257 (53) & 304 (56)\\
\hline
\rule[-1ex]{0pt}{3.5ex}  $\kappa$~(nW/$\mathrm{K^4}$) & 16.1 (3.0) & 16.9 (1.0) & 13.9 (3.0) & 15.2 (0.9)\\
\hline
\rule[-1ex]{0pt}{3.5ex}  $P_{\mathrm{sat}}$~(pW) @ 50~mK & 10.7 (2.9) & 8.4 (1.5) & 10.7 (2.6) & 12.7 (3.1)\\
\hline
\rule[-1ex]{0pt}{3.5ex}  $R_\mathrm{N}$~($\mathrm{m\Omega}$) & 12.7 (0.5) & 11.0 (0.5) & 10.3 (0.4) & 10.7 (0.3) \\
\hline
\rule[-1ex]{0pt}{3.5ex}  $T_\mathrm{c}$~(mK) & 163 (9) & 151 (8) & 165 (10) & 171 (10)\\
\hline
\end{tabular}
\end{center}
\end{table}

\begin{figure} [ht]
   \begin{center}
   \includegraphics[width=\textwidth,trim={1.25cm 0 0 0}]{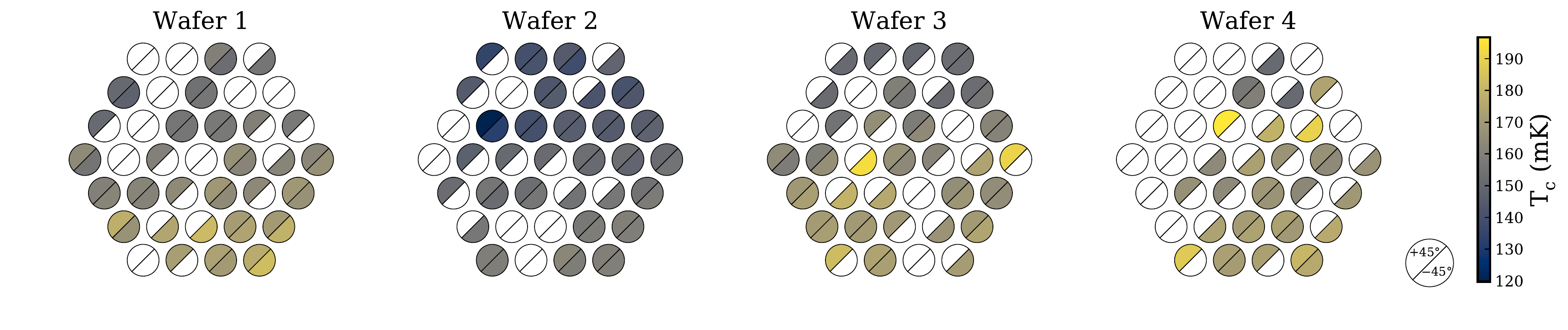}
   \end{center}
   \caption{ \label{fig:Tcwaferplot} 
Measured values of $T_\mathrm{c}$ at each detector pixel. $T_\mathrm{c}$  values shown are for 175 bolometers that produced good I-V measurements during in-lab dark testing at JHU. The circles are split to show both orthogonal polarization states ($\pm45\degree$) contained within a pixel. Median values for each wafer are shown in Table~\ref{tab:parameters}.}
   \end{figure}

In Table~\ref{tab:parameters}, we report the median derived parameters and their standard deviations across each of the four new 90~GHz wafers. The reported values are calculated using approximately 45, 51, 47, and 32 bolometers, respectively.  Using the fitted $T_\mathrm{c}$ and $\kappa$ values, we report the calculated $P_{\mathrm{sat}}$ at our operating $T_{\mathrm{bath}}$ of $\sim$~50~mK.  Figure~\ref{fig:Tcwaferplot} shows the measured values of $T_\mathrm{c}$ at each detector pixel.

\subsection{Bandpass Measurements}
\label{subsec:bandpass_measurements}

Bandpass measurements were performed using a Fourier transform spectrometer (FTS), which was built at JHU.\cite{FTS}
The FTS is a Martin–Puplett interferometer with a movable mirror that scans back and forth with a range of 150~mm at 0.5~mm/s.
The input signal of the FTS is a wide-band thermal source, at about $1000\degree\mathrm{C}$.
The output signal of the FTS, measured by the detectors, is modulated with a chopper at 21~Hz.
The FTS has a resolution of $\sim$1.6~GHz.
The bandpass is given by the real component of the Fourier transform of the interferogram.
The measured response plotted in Figure~\ref{fig:bandpass} is the inverse variance-weighted average over 93 bolometers that yielded high quality interferograms and bandpasses across three of the four new 90~GHz wafers.
The measured bandpass edges are in good agreement with the simulations; the discrepancies observed in-band between measured and simulated responses are likely due to optical effects related to the test setup that are not included in the simulation.
In Table~\ref{tab:bandpass}, we report the measured bandwidths and effective center frequencies associated with various types of diffuse sources following the methods described in \S~3.1 of Ref.~\citenum{Dahal-multifrequency}.
Uncertainties for the bandwidths (effective center frequencies) are given by the summation in quadrature of the standard errors on the mean and the FTS measurement resolution (half FTS measurement resolution).

\newcolumntype{P}[1]{>{\centering\arraybackslash}m{#1}}

   \begin{minipage}{\linewidth}
      \begin{minipage}{0.35\linewidth}
          \begin{figure}[H]
   \includegraphics[width=\linewidth, trim={1.25cm 0 0 0}]{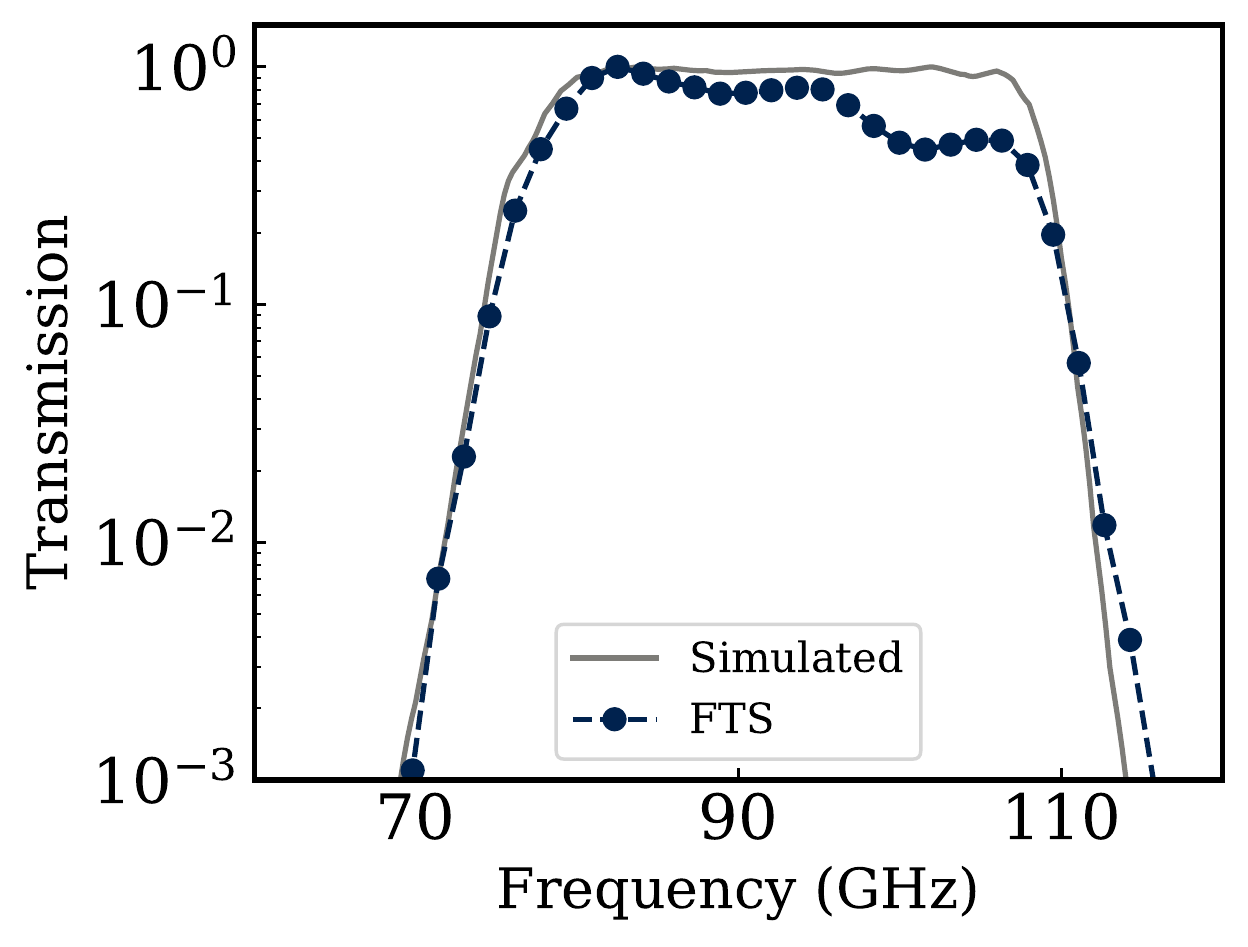}
              \caption{\label{fig:bandpass}Simulated and inverse-variance weighted average measured spectral response (normalized to unity) for 93 bolometers across three wafers. The bandpass edges are in agreement with simulations.}
          \end{figure}
      \end{minipage}%
      \hspace{0.4cm}
      \begin{minipage}{0.59\linewidth}
          \begin{table}[H]
            \caption{Measured Bandwidths and Effective Center Frequencies for various types of diffuse sources.}
            \label{tab:bandpass}
            \begin{center}       
            \begin{table}[H]
            \begin{tabular}{cP{1.6cm}l|c|P{2.5cm}|}
            
            \cline{1-2} \cline{4-5}
            \multicolumn{1}{|c|}{\rule[-1ex]{0pt}{3.5ex}} & \multicolumn{1}{P{1.6cm}|}{Bandpass (GHz)} &  & Source & Effective Center Frequencies (GHz) \\ \cline{1-2} \cline{4-5} 
            \multicolumn{1}{|c|}{\rule[-1ex]{0pt}{3.5ex}FWHP}  & \multicolumn{1}{P{1.6cm}|}{21.5±1.8} &  & Synchrotron    & 88.9$\pm$0.92 \\ \cline{1-2} \cline{4-5} 
            \multicolumn{1}{|c|}{\rule[-1ex]{0pt}{3.5ex}Dicke} & \multicolumn{1}{P{1.6cm}|}{32.0±1.7} &  & Rayleigh–Jeans & 91.2$\pm$0.92 \\ \cline{1-2} \cline{4-5} 
                                        \rule[-1ex]{0pt}{3.5ex} &                               &  & Dust           & 92.7$\pm$0.92 \\ \cline{4-5} 
                                        \rule[-1ex]{0pt}{3.5ex} &                               &  & CMB            & 90.8$\pm$0.92 \\ \cline{4-5} 
            \end{tabular}
            \end{table}
            \end{center}
        \end{table}
        
      \end{minipage}%
  \end{minipage}

\subsection{Noise-Equivalent Power}
\label{subsec:noise_performance}

\begin{figure} [ht]
   \begin{center}
   \begin{tabular}{c} 
   \includegraphics[width=\linewidth]{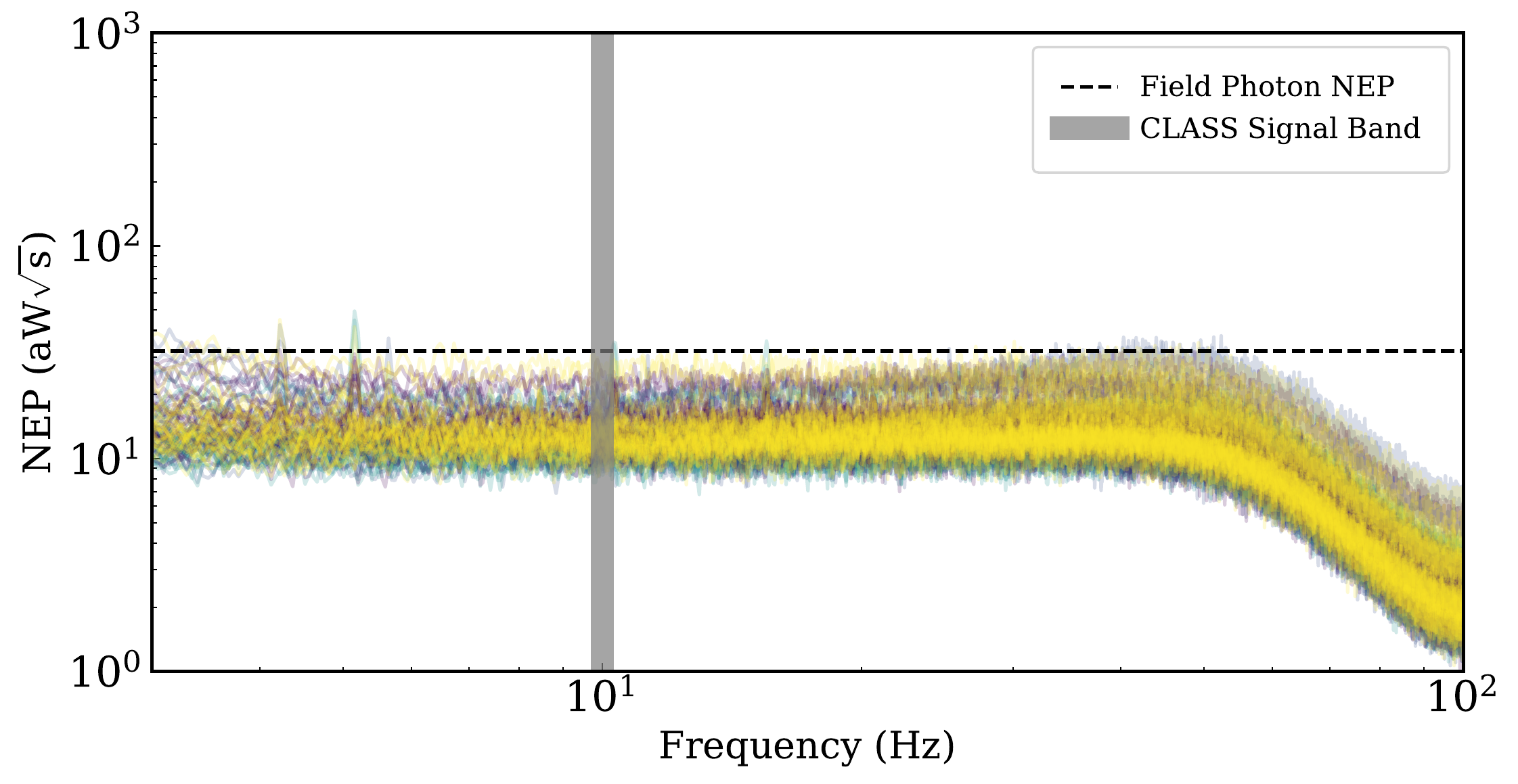}
   \end{tabular}
   \end{center}
   \caption[example] 
   { \label{fig:NEP} 
Measured dark NEP for 106 TES bolometers.
Individual detectors are plotted, with line color (purple, blue, green, yellow) corresponding to each of the four wafers. The gray vertical line shows the modulated CLASS signal band.\cite{Katie-VPM} The dashed horizontal line shows the expected photon NEP at the CLASS site due to the CMB as well as instrumental and atmospheric emission.
The measured detector NEP during dark tests is below the expected NEP of 32 $\mathrm{aW\sqrt{s}}$ from photon noise in the field; therefore we expect the new 90~GHz detectors to be photon noise limited.}
   \end{figure}

We estimate the noise performance of the new detectors by measuring the NEP from dark tests for all four new 90~GHz modules.
The resulting noise spectra for 106 TES bolometers are shown in Figure~\ref{fig:NEP}.
We find a median NEP of 12.3 $\mathrm{aW\sqrt{s}}$ in the 8--12~Hz window about the CLASS modulation frequency of 10~Hz above the $1/f$ instrumental and atmospheric noise.\cite{Katie-VPM}
The observed roll-off at high frequencies is due to the MCE digital Butterworth filter that is applied to suppress noise aliasing from higher frequencies.
We anticipate the CLASS 90~GHz detectors to be photon noise limited.
The measured dark NEP is below the expected photon NEP of 32 $\mathrm{aW\sqrt{s}}$ in the field due to emission from the CMB, the atmosphere, and the instrument.

\section{Conclusion}
\label{sec:conclusion}

We present preliminary characterization, via electrothermal parameters, bandpass measurements, and dark noise measurements, of four new CLASS 90~GHz detector wafers.
The wafers include an updated TES architecture that aims to improve the stability and optical efficiency of the detectors.
These detectors have been installed in the field, upgrading four modules of the W1 focal plane.
The new detectors are anticipated to achieve first light in 2022.
Further in-lab and on-sky characterization will be presented in a future publication.

\acknowledgments 

We acknowledge the National Science Foundation Division of Astronomical Sciences for their support of CLASS under Grant Numbers 0959349, 1429236, 1636634, 1654494, 2034400, and 2109311.
We thank Johns Hopkins University President R. Daniels and the Krieger School of Arts and Sciences Deans for their steadfast support of CLASS.
We further acknowledge the very generous support of Jim and Heather Murren (JHU A\&S '88), Matthew Polk (JHU A\&S Physics BS '71), David Nicholson, and Michael Bloomberg (JHU Engineering '64).
The CLASS project employs detector technology developed in collaboration between JHU and Goddard Space Flight Center under several previous and ongoing NASA grants. Detector development work at JHU was funded by NASA cooperative agreement 80NSSC19M0005.
Kyle Helson is supported by NASA under award number 80GSFC17M0002.
Zhilei Xu is supported by the Gordon and Betty Moore Foundation through grant GBMF5215 to the Massachusetts Institute of Technology.
We acknowledge scientific and engineering contributions from Max Abitbol, Fletcher Boone, David Carcamo, Lance Corbett, Ted Grunberg, Saianeesh Haridas, Jake Hassan, Connor Henley, Ben Keller, Lindsay Lowry, Nick Mehrle, Sasha Novak, Diva Parekh, Isu Ravi, Gary Rhodes, Daniel Swartz, Bingjie Wang, Qinan Wang, Tiffany Wei, Zi\'ang Yan, and Zhuo Zhang. We thank Miguel Angel, Jill Hanson, William Deysher, Mar\'ia Jos\'e Amaral, and Chantal Boisvert for logistical support.
We acknowledge productive collaboration with Dean Carpenter and the JHU Physical Sciences Machine Shop team.
CLASS is located in the Parque Astron\'omico Atacama in northern Chile under the auspices of the Agencia Nacional de Investigaci\'on y Desarrollo (ANID).

\bibliography{main} 
\bibliographystyle{spiebib} 

\end{document}